\documentclass[prl,12pt,draft,amsmath,amssymb,showpacs]{revtex4}

\begin{document}

\title{Fine structure constant variation or spacetime anisotropy?}

\author{Zhe Chang}
\email{changz@ihep.ac.cn}
\author{Sai Wang}
\email{wangsai@ihep.ac.cn}
\author{Xin Li}
\email{lixin@ihep.ac.cn}

\affiliation{Institute of High Energy Physics\\
Theoretical Physics Center for Science Facilities\\
Chinese Academy of Sciences, 100049 Beijing, China}


\begin{abstract}
Recent observations on the quasar absorption spectra supply evidence for the variation of the fine structure constant $\alpha$. In this paper, we propose another interpretation of the observational data on the quasar absorption spectra: a scenario with spacetime inhomogeneity and anisotropy.
Maybe the spacetime is characterized by the Finsler geometry instead of the Riemann one. The Finsler geometry admits fewer symmetries than the Riemann geometry does. We investigate the Finslerian geodesic equations in the Randers spacetime (a special Finsler spacetime). It is found that the cosmological redshift in this spacetime deviates from the one in general relativity. The modification term to the redshift could be generally revealed as a monopole plus dipole function of spacetime locations and directions. We suggest that this modification corresponds to the spatial monopole and dipole of $\alpha$ variation in the quasar absorption spectra.
\end{abstract}

\maketitle

\section{I. Introduction}

It has been widely accepted that the standard cosmological model ($\Lambda$CDM model) \cite{Book by Dodelson}
is the paradigm of the modern cosmology.
This model makes several observable predictions
which have withstood large quantities of tests by the cosmological observations during the last two decades.
Until now, almost all observations, such as the cosmic microwave background (CMB) anisotropy \cite{KomatsuETAL01},
the cosmological accelerating expansion \cite{KesslerETAL01,AmanullahETAL01}
and the large scale structure (LSS) \cite{NesserisP01}, agree well with the predictions of $\Lambda$CDM model.
Thus, this model is indeed great successful.
Despite these, it still faces several cosmological large scale anomalies (see review in Ref. \cite{AntoniouP01}),
such as the large scale velocity flows \cite{WatkinsFH001}, the alignment of the low multipoles in the CMB spectra \cite{BennettETAL001,LandM001,TegmarkOH001},
the large scale alignment in the quasar optical polarization \cite{HutsemekersCLS001}
and the preferred axis of Hubble diagram \cite{AntoniouP01,SchwarzW001}.
These anomalies imply that there may exist inhomogeneity and anisotropy at large scales
which are introduced by certain common preferred direction in the spacetime \cite{AntoniouP01}.
All of these are beyond the $\Lambda$CDM model and may lead to new physics.

It is well known that there are in principle no variations of the fundamental physical constants,
such as the fine structure constant \(\alpha={e^{2}}/{\hbar c}\)
($e$ is the unit electric charge, $\hbar$ is the reducible Planck constant and $c$ is the speed of light in vacuum),
in the $\Lambda$CDM model which is based on the cosmological principle \cite{Weinberg001} and Einstein's general relativity \cite{Weinberg002}.
However, there have been several propositions for the existence of possible $\alpha$ variation currently.
Since Dirac \cite{Dirac001} postulated in 1937 that the universal gravitational constant $G$ is not a constant possibly,
great interests have been stimulated in studies on the variation of fundamental physical constants including the fine structure constant $\alpha$.
A variety of experiments and observations \cite{UzanJP001,UzanJP002} has been employed in searching for the $\alpha$ variation.
Meanwhile, quantities of theories or models (see review and details in Ref.\cite{UzanJP002}) have been proposed and studied,
which suggest various possibilities for the $\alpha$ variation.
There are several reasons for the possible $\alpha$ variation (see review and details in Ref.\cite{UzanJP002,BarrowJD001}),
such as the existence of the extra dimensions, quantum gravity, the nonuniqueness of vacuum state,
the spontaneous symmetry breaking in the very early universe and \emph{etc}.
Thus, it is interesting and meaningful to study the possible $\alpha$ variation.

To search for the $\alpha$ variation, there are three classes of experiments:
atomic methods, nuclear methods, and gravitational methods (for review and details see, for example, Ref.\cite{UzanJP001}).
Recently, astrophysical observations provided some evidence for the $\alpha$ variation \cite{MurphyFW01,BerengutF01,WebbKMFCB01}.
The many-multiplet (MM) method \cite{DzubaFW01,WebbFCCMDB01,DzubaFW02}
was employed to analyze the data of quasar absorption spectra \cite{DzubaFW01,VarshalovichP01}.
This method utilizes the measurement of the difference between two wavelengths and comparisons with the laboratorial value
to determine the $\alpha$ variation \cite{Flambaum01,BerengutF01}.
In addition, the quasar absorption spectra encode information of the atomic transitions at the positions and moments of emissions \cite{UzanJP001}, so that analysis of them will reveal information about $\alpha$ around the distant quasars.
One has to determine the $\alpha$-dependence of the atomic spectra in order to observe the $\alpha$ variation around the distant quasars \cite{UzanJP001}.
In the condition that $\alpha$ has a small shift (\({\delta\alpha}/{\alpha}=\frac{\alpha-\alpha_{0}}{\alpha_{0}}\ll1\)),
the transitions between the fine structure multiplets are described by \cite{UzanJP001,MurphyFW01}
\begin{equation}
\label{omega}
\omega=\omega_{0}+q x\equiv \omega_{0}+q\cdot\frac{2 \delta\alpha}{\alpha}\ ,
\end{equation}
where \(x\equiv{2\delta\alpha}/{\alpha}\).
Throughout the paper, $\omega$ and $\alpha$ also represent, respectively, the atomic transitions
and the fine structure constant at the positions and moments of the emissions from the quasars with the redshift \(z={\lambda_{obs}}/{\lambda_{lab}}-1\).

Recently, Webb \emph{et al.} \cite{MurphyFW01} analyzed the data of quasar absorption spectra
from the Keck--Hires Telescope (Keck) with the MM method.
They claimed that $\alpha$ is smaller at large scales
\begin{equation}
\label{monopole}
\frac{\delta\alpha}{\alpha}(z)=(-0.543\pm0.116)\times10^{-5}\ .
\end{equation}
This is a spatial monopole function about redshift $z$.
Most recently, they \cite{BerengutF01,WebbKMFCB01} claimed that evidence for a spatial dipole, named as ``Australian Dipole'',
of the $\alpha$ variation is found also by the quasar absorption spectra from the Keck, the Very Large Telescope (VLT) and both combined.
They found that $\alpha$ varies in the form of a spatial dipole which is larger at one hemisphere and smaller at the other one at large scales.
The spatial dipole of the $\alpha$ variation was revealed as \cite{WebbKMFCB01}
\begin{equation}
\label{dipole}
\frac{\delta\alpha}{\alpha}\left(\cos\Theta, z\right)=(1.10\pm0.25)\times10^{-6}\ ,
\end{equation}
where $\Theta$ is the angle between the quasar sightline and the best-fit dipole position.
The spatial monopole and dipole of the $\alpha$ variation were found to take the same order of magnitude $\sim10^{-6}$.

It is well known that Glashow and Cohen proposed the very special relativity (VSR) \cite{CohenG01}
in which the Lorentz group is replaced by its subgroup.
There exists a preferred direction \cite{GibbonsGP01,Bogoslovsky02,Bogoslovsky03,KouretsisSS01} in VSR,
which leads to the Lorentz invariance violation (LIV).
It has also been clear that the preferred direction could give rise to anisotropy of the speed of light
in the vacuum \cite{ColemanG001,ColemanG002}.
Thus the fine structure constant $\alpha$ would vary with directions and show anisotropy in space,
since it has an inverse ratio dependency on the speed of light.
In addition, the line element of VSR has been proved to be a Finslerian line element \cite{GibbonsGP01,Bogoslovsky02,Bogoslovsky03,KouretsisSS01}.
Therefore, Finsler spacetime could bring about new insights on the $\alpha$ variation.
Finsler spacetime admits less Killing vectors (equivalently fewer symmetries) than the Riemann one does \cite{ChangL06}.
This means that there exist preferred directions which lead to the inhomogeneity and anisotropy in the Finsler spacetime.
Such inhomogeneity and anisotropy would lead to the $\alpha$ variation with locations and directions in the spacetime,
which are the LIV effects that make the Finsler spacetime different from the $\Lambda$CDM model.
The attribute of the spatial monopole of the $\alpha$ variation implies that the universe is inhomogeneous
and the existence of the spatial dipole implies that there may exist anisotropy at large scales.
As mentioned above, this kind of anisotropy may be also the reasons why other cosmological large scale anomalies emerge \cite{AntoniouP01}.
The inhomogeneity and anisotropy of this kind signal certain nontrivial cosmic topology \cite{LandM001}.
Then there may exist a special kind of spacetime structure at large scales
other than the usual Friedmann--Robertson--Walker (FRW) structure in the $\Lambda$CDM model.
Maybe the Finsler spacetime is a reasonable candidate for new physics correspond to the $\alpha$ variation claimed.
These are new results in this work and may potentially lead to new physics.

In Einstein's general relativity, gravity is connected with the curvature in the Riemann geometry.
In the same way, one could discuss gravity based on the Finsler geometry \cite{Book by Bao,Book by Shen}.
Gravity in the Finsler spacetime has been studied for a long time \cite{Takano01,Ikeda01,Tavakol01,Bogoslovsky01}.
An incomplete \emph{list} of works in this field includes studies of: a specified Finsler structure makes modified Newton's gravity \cite{ChangL01} equivalent to Milgrom's Modified Newtonian Dynamics (MOND) \cite{Milgrom01}; a Finlerian gravity model accounts for the accelerated expanding universe without invoking the dark energy hypothesis \cite{ChangL02}; the Randers space \cite{ChangL03} accounts for the anomalous acceleration \cite{Anderson01} in the solar system observed by Pioneer $10$ and $11$ spacecrafts; the Finsler spacetime leads to modification of the gravitational deflection of light \cite{ChangL04} corresponding to observations on the Bullet Cluster \cite{CloweRM01}; the Finslerian kinematics is in good agreement with the secular trend of the Astronomical Unit and the secular eccentricity variation of the Moon's orbit \cite{ChangL05};
the Finsler--Schwarzchild metric asymptotically approaches the Bogoslovsky locally anisotropic spacetime
instead of the Minkowski spacetime \cite{Silagadze01}.

In this paper, we suggest an inhomogeneous and anisotropic spacetime could describe well the astronomical observations on the quasar absorption spectra. The rest of the paper is arranged as follows. In section II, we discuss the spacetime inhomogeneity and anisotropy in the framework of the Finsler geometry. A uniform formula for the cosmological redshift is presented in section III. The observed spatial monopole and Australian Dipole is fitted in the new scenario. We give the conclusions and remarks in section IV.

\section{II. Spacetime inhomogeneity and anisotropy}

The Finsler geometry \cite{Book by Bao,Book by Shen} originates from the integrals of form
\begin{eqnarray}
\int^b_a F\left(x, y\right)d\tau\ ,
\label{integral length}
\end{eqnarray}
where $x$ and $y \equiv dx/d\tau$ stand, respectively, for the position and the velocity under the natural coordinate bases.
The integrand $F$ is called a Finsler structure.
Unlike the Riemann structure being defined on the manifold $M$, the Finsler structure is defined on the slit tangent bundle \{$TM\setminus0$\}.
A Finsler structure of $M$ is a positive-definite function with the property
\begin{equation}
F(x,\lambda y)=\lambda F(x,y)
\end{equation}
for all $\lambda>0$.
A manifold $M$ associated with a Finsler structure $F$ on \{$TM\setminus0$\} is called a Finsler manifold.
The Finsler metric tensor is a Hessian matrix, the coefficients of which are defined as \cite{Book by Bao}
\begin{equation}
g_{\mu\nu}\equiv\frac{\partial}{\partial y^\mu}\frac{\partial}{\partial y^\nu}\left(\frac{1}{2}F^2\right).
\end{equation}
It is also called the fundamental tensor and is used for raising and lowering the indices together with their inverse $g^{\mu\nu}$.

The parallel transport has been studied in the framework of Cartan connection \cite{Matsumoto01,Antonelli01,Szabo01}.
The notation of parallel transport in the Finsler manifold means that the length \(F\left(\frac{dx}{d\tau}\right)\) is constant.
The geodesic equation in the Finsler manifold is given as \cite{Book by Bao}
\begin{equation}
\label{geodesic}
\frac{d^2x^\mu}{d\tau^2}+G^\mu=0\ ,
\end{equation}
where
\begin{equation}
\label{geodesic spray}
G^\mu=\frac{1}{2}g^{\mu\nu}\left(\frac{\partial^2 F^2}{\partial x^\lambda \partial y^\nu}y^\lambda-\frac{\partial F^2}{\partial x^\nu}\right)
\end{equation} are the geodesic spray coefficients.
Obviously, if $F$ is the Riemannian metric, then
\begin{equation}
G^\mu=\tilde{\gamma}^\mu_{\nu\lambda}y^\nu y^\lambda\ ,
\end{equation}
where $\tilde{\gamma}^\mu_{\nu\lambda}$ is the Riemannian Christoffel symbol.
Since the geodesic equation (\ref{geodesic}) is directly derived from the integral length of a curve $\sigma$
\begin{equation}
L(\sigma)=\int F\left(\frac{dx}{d\tau}\right)d\tau\ ,
\end{equation}
the inner product \(\left(\sqrt{g_{\mu\nu}\frac{dx^\mu}{d\tau}\frac{dx^\nu}{d\tau}}=F\left(\frac{dx}{d\tau}\right)\right)\)
of two parallel transported vectors is preserved.

The Randers space \cite{Randers space} is a special kind of Finsler space with the Finsler structure $F$ of the form
\begin{equation}
F(x,y)\equiv \alpha(x,y)+\beta(x,y)\ ,
\end{equation} where
\begin{eqnarray}
\alpha (x,y)&\equiv&\sqrt{\tilde{a}_{\mu\nu}(x)y^\mu y^\nu}\ ,\\
\beta(x,y)&\equiv& \tilde{b}_\mu(x)y^\mu,
\end{eqnarray} and $\tilde{a}_{ij}$ is the Riemannian metric.

The geodesic spray coefficient $G^\mu$ in the Randers--Finsler spacetime reads \cite{Book by Bao}
\begin{eqnarray}
G^\mu&=&(\tilde{\gamma}^\mu_{\nu\lambda}+l^\mu\tilde{b}_{\nu|\lambda})y^\nu y^\lambda\nonumber\\
&+&(\tilde{a}^{\mu\nu}-l^\mu\tilde{b}^\nu)(\tilde{b}_{\nu|\lambda}-\tilde{b}_{\lambda|\nu})\alpha\left(\frac{dx}{d\tau}\right) y^\lambda\ ,
\end{eqnarray}
where $l^\mu\equiv y^\mu/F$, $\tilde{\gamma}^\mu_{\nu\lambda}$ is the Christoffel symbol of the Riemannian metric $\tilde{a}$ and $\tilde{b}_{\nu|\lambda}$ denotes the covariant derivative with respect to the Riemannian metric $\tilde{a}$
\begin{equation}
\tilde{b}_{\nu|\lambda}=\frac{\partial\tilde{b}_\nu}{\partial x^\lambda}-\tilde{\gamma}^\mu_{\nu\lambda}\tilde{b}_\mu\ .
\end{equation}
In the rest of the paper, we just consider the case that $\beta$ is a closed 1-form.
Thus, the geodesic equation of such a Randers spacetime is given as
\begin{equation}
\label{geodesic randers}
\frac{d^2x^\mu}{d\tau^2}+(\tilde{\gamma}^\mu_{\nu\lambda}+l^\mu\tilde{b}_{\nu|\lambda})y^\nu y^\lambda=0\ .
\end{equation}

\vspace{0.5cm}

In the $\Lambda$CDM model, the cosmological principle indicates that the universe is homogeneous and isotropic at large scales,
which leads to the Friedmann--Robertson--Walker (FRW) metric.
In comoving coordinates, the FRW metric takes the form
\begin{equation}
ds^2=dt^2-a^2(t)\left[\frac{dr^2}{1-kr^2}+r^2\left(d\theta^2+\sin^2\theta d\varphi^2\right)\right] ,
\end{equation}
where \(k=-1, 0, +1\), respectively, stand for open, flat, and closed universe, and $a(t)$ is the scale factor.
The cosmological redshift $z_{R}(t)$ is given as
\begin{equation}
\label{redshiftinFriemann}
1+z_{R}(t)=\frac{a(t_{0})}{a(t)}=\frac{1}{a}\ ,
\end{equation}
which reveals the ratio of expansion undergone by the universe between time $t$ and present $t_{0}$.

Unfortunately, the FRW universe does not match the Keck and VLT observations.
The attribute of the spatial monopole in $\alpha$ variation implies that the universe is not homogeneous,
and the ``Australian Dipole ''implies that the universe is not isotropic.
As is mentioned above, the Finsler spacetime naturally admits less Killing vectors (therefore fewer symmetries) than the Riemann one does.
It could be a reasonable framework to incorporate the Keck and VLT observations.

\section{III. Uniform redshift and Australian dipole}

We suppose that the metric of the universe takes the FRW--Randers--Finsler form,
in which $\tilde{a}_{\mu\nu}$ is the flat FRW metric and $\beta$ is a closed 1-form.
Then, we find from (\ref{geodesic randers}) that
\begin{eqnarray}
\label{geodesic eq1}
0&=&\frac{d^2x^0}{d\tau^2}+\delta_{ij}\dot{a}a\frac{dx^i}{d\tau}\frac{dx^j}{d\tau}+\frac{d x^{0}}{d\tau}f\left(x,\frac{dx}{d\tau}\right),\\
\label{geodesic eq2}
0&=&\frac{d^2x^i}{d\tau^2}+2\delta^i_j\frac{\dot{a}}{a}\frac{dx^0}{d\tau}\frac{dx^j}{d\tau}+\frac{dx^i}{d\tau}f\left(x,\frac{dx}{d\tau}\right),
\end{eqnarray}
where \(f\left(x,\frac{dx}{d\tau}\right)\equiv\tilde{b}_{\nu|\lambda}\frac{dx^\nu}{d\tau}\frac{dx^\nu}{d\tau}/F\) and a dot denotes $\frac{d}{dx^0}$.
Equation (\ref{geodesic eq2}) has a solution
\begin{equation}\label{geodesic so1}
a^2\frac{dx^i}{d\tau}\propto J_1,
\end{equation}
involving a new quantity $J_1$. It is defined as \(\frac{d\ln J_1}{d\tau}\equiv-f\left(x,\frac{dx}{d\tau}\right)\). By making use of equation (\ref{geodesic so1}), we obtain a solution of (\ref{geodesic eq1}),
\begin{equation}
\label{geodesic so2}
a\frac{dx^0}{d\tau}\propto J_1~.
\end{equation}
While $\tilde{b}$ vanishes, $J_1$ reduces to a dimensionless constant for photons.
The Riemannian norm $\tilde{b}$ is much smaller than 1.
Therefore, the energy of the universe is of its Riemannian form \(E\simeq\frac{dx^0}{d\tau}\).
Then, we find from the solution (\ref{geodesic so2}) that the formula of redshift in the FRW--Rander--Finsler spacetime is of the form
\begin{equation}
\label{Fi}
1+z_F(t)=\frac{J_1}{a}\ .
\end{equation}

In the following, we try to get a formula for $J_1$ . The derivative of the term \(\tilde{b}_\mu\frac{dx^\mu}{d\tau}\) gives
\begin{eqnarray}
\frac{d}{d\tau}\left(\tilde{b}_\mu\frac{dx^\mu}{d\tau}\right)&=&\frac{dx^\nu}{d\tau}\frac{\partial}{\partial x^\nu}\left(\tilde{b}_\mu\frac{dx^\mu}{d\tau}\right)=\frac{dx^\nu}{d\tau}\left(\tilde{b}_\mu\frac{dx^\mu}{d\tau}\right)_{|\nu}\nonumber\\
&=&\tilde{b}_{\alpha|\beta}\frac{dx^\alpha}{d\tau}\frac{dx^\beta}{d\tau}\nonumber\\
&+&\tilde{b}_\mu\left(\frac{d^2x^\mu}{d\tau^2}+
\tilde{\gamma}^\mu_{\nu\lambda}\right)\frac{dx^\nu}{d\tau}\frac{dx^\nu}{d\tau}\nonumber\\
\label{deduce J1}
&=&\left(1-\frac{\tilde{b}_\mu}{F}\frac{dx^\mu}{d\tau}\right)\tilde{b}_{\alpha|\beta}\frac{dx^\alpha}{d\tau}\frac{dx^\beta}{d\tau}\ ,
\end{eqnarray}
where $``|"$ denotes the covariant derivative with respect to the Riemannian metric $\alpha$.
Here, we have used the fact that the term \(\tilde{b}_\mu\frac{dx^\mu}{d\tau}\) is a scaler in the Riemannian spacetime with metric $\tilde{a}_{\mu\nu}$, to get the second equation of (\ref{deduce J1}).
Also we have used the geodesic equation (\ref{geodesic randers}) to get the last equation of (\ref{deduce J1}).
Noticing that $F$ is constant along the geodesic, we find from equation (\ref{deduce J1}) that
\begin{eqnarray}
\frac{d\ln\left(F-\tilde{b}_\mu\frac{dx^\mu}{d\tau}\right)}{d\tau}&=&-\tilde{b}_{\nu|\lambda}\frac{dx^\nu}{d\tau}\frac{dx^\nu}{d\tau}/F\nonumber\\
&=&-f\left(x,\frac{dx}{d\tau}\right).
\end{eqnarray}
It implies that
\begin{equation}
\label{formula J1}
J_1=1-\tilde{b}_\mu\frac{dx^\mu}{d\tau},
\end{equation}
with normalization of $\tau$ ($F$ has been normalized).

Combining equations (\ref{redshiftinFriemann}), (\ref{Fi}) and (\ref{formula J1}) together,
we obtain the cosmological redshift deviation in the Finsler spacetime from the one in the Riemann spacetime
\begin{equation}
\label{modifiedredshift}
\frac{1+z_{R}}{1+z_{F}}\simeq1+\tilde{b}_{\mu}\hat{p}^{\mu}\ ,
\end{equation}
where the over-hat represents that $\hat{p}^{\mu}$ is a unit four-momentum of light.
Here, we have assumed that there is no $\alpha$ variation around the Earth, which is implied by the ``ether drift'' experiment mentioned below.

It is obvious that the second term, \(\tilde{b}_{\mu}\hat{p}^{\mu}\), on the right hand side of equation (27) could be rewritten into a monopole plus dipole function about spacetime locations and directions.
The dipole term (proportional to \(\cos\Theta:=\hat{b}_{i}\cdot\hat{p}^{i}\))
comes from the inner product of the space components of $\tilde{b}^{\mu}$ and $\hat{p}^{\mu}$,
and the monopole one (which is (\(\tilde{b}_{0}\hat{p}^{0}\))) comes from the inner product of the time components of them.
We may live in an inhomogeneous and anisotropic universe, but used to calculate all physical quantities from a homogeneous and isotropic viewpoint.
This may be the reason why the observational fine structure constant $\alpha$ varies from point to point in the spacetime.
Thus, one should try to setup a new scenario of the observational cosmology with the spacetime inhomogeneity and anisotropy, and deal with the observational data of the quasar absorption spectra from a uniform view of the FRW--Randers--Finsler spacetime.
In the frame of the Finsler spacetime, the relative frequency $\Delta\omega$ ($\Delta$ means the relative value between two parameters) between two given transitions emitted from the distant quasars should be observed or detected as \((\Delta\omega_{0})/(1+z_{F})\) on the Earth.
However, the observational data from Keck and VLT showed that it takes the form \(((\Delta\omega_{0})+(\Delta q) x)/(1+z_{R})\)
in the perspective of the FRW--Riemann spacetime.
The relative frequency of the two given transitions predicted by the Finsler spacetime
should equal the observational relative frequency of them by Keck and VLT from the viewpoint of the new scenario of the observational cosmology
\begin{equation}
\label{Finsler=observations}
\frac{\Delta\omega_{0}}{1+z_{F}}=\frac{\Delta\omega_{0}+\Delta q\cdot \frac{2 \delta\alpha}{\alpha}}{1+z_{R}}\ ,
\end{equation}
where we have substituted $x$ with \(2\delta\alpha/\alpha\).
By combining equations (\ref{omega}), (\ref{modifiedredshift}) and (\ref{Finsler=observations}) together,
we could obtain the claimed formulas for the spatial monopole (\ref{monopole}) and the spatial dipole (\ref{dipole}) of
the $\alpha$ variation observed by Keck and VLT,
\begin{equation}
\label{modifiedredshift1}
\frac{\delta\alpha}{\alpha}=A+B\cos\Theta\ ,
\end{equation}
where \(A=\frac{\Delta\omega_{0}}{2\Delta q}(\tilde{b}_{0}\hat{p}^{0})\) and \(B=\frac{\Delta\omega_{0}}{2\Delta q}|\tilde{b}^{i}|\).
Both the dipole and monopole terms of the $\alpha$ variation appear naturally in the Finsler spacetime.
Under the condition that both the monopole $A$ and dipole $B$ terms in equation (\ref{modifiedredshift1})
take the order of magnitude $\sim10^{-7}$, the $\alpha$ variation would appear to be of the order \(\sim10^{-6}\),
which is consistent with the formulas of the $\alpha$ variation
from the Keck and VLT observations on the quasar spectra with redshift \(0.2\sim4.2\).
Of course, if the Randers--Finsler metric holds at the cosmological scale, it could account for the accelerating expansion of the universe \cite{ChangLL008}.

\section{IV. Conclusions and remarks}

The line element of VSR reads \cite{GibbonsGP01,Bogoslovsky02,Bogoslovsky03,KouretsisSS01}:
\begin{equation}
\label{VSR}
ds=\left(\eta_{\mu\nu}y^{\mu}y^{\nu}\right)^{\frac{1-b}{2}}\left(n_{\sigma}y^{\sigma}\right)^{b}d\tau\ ,
\end{equation}
where the unit vector $n_{k}$ denotes the preferred direction in the three-dimensional space.
The tiny parameter $b$ stands for the level of spacetime anisotropy, which characterizes the deviation of the VSR metric (\ref{VSR})
from the Minkowski one \(\eta_{\mu\nu}=diag(+1,-1,-1,-1)\).
In the case that \(\left(\left(n_{\sigma}y^{\sigma}\right)/\left(\eta_{\mu\nu}y^{\mu}y^{\nu}\right)^{1/2}-1\right)\)
is small, the right hand side of equation (\ref{VSR}) could be expanded into the form of
\begin{equation}
ds\approx\left((1-b)\sqrt{\eta_{\mu\nu}y^{\mu}y^{\nu}}+bn_{\sigma}y^{\sigma}\right)d\tau\ .\\
\end{equation}
It is obvious that this is a Randers line element.
Furthermore, studies on the Killing vectors of VSR show that there exists anisotropy in the spacetime \cite{ChangL06}.
Thus, the experimental tests of the space anisotropy may provide constraints on the Randers spacetime.
From the ``ether drift'' experiment obtained in 1970 \cite{Bogoslovsky02,ChampeneyIK01,Isaak01},
the space anisotropy has an upper limit \(b<5\times10^{-10}\).
This means that there exist no significant preferred directions around the Earth.
In fact, we have already used the assumption that \(\tilde{b}_{\mu}(z=0)=0\) to derive equation (\ref{modifiedredshift}).
The observations of the quasar spectra by the Keck and VLT hinted that there are effects of a preferred direction
at the places with redshift \(0.2\sim4.2\) \cite{WebbKMFCB01}.

The reports of the evidence for the $\alpha$ variation from the observations by the Keck and the VLT may signal prelude of the new physics
if it was confirmed by other independent observations or experiments.
However, the reports of the $\alpha$ variation currently do not provide statistically significant evidence
for the inhomogeneity and anisotropy of the cosmology.
The reasons include: there are only $153$ data in the observations of the $\alpha$ variation and the observed quasars by Keck and VLT spread
in two separate skies, respectively, at two opposite hemispheres \cite{WebbKMFCB01}.
However, the $\Lambda$CDM model has withstood large numbers of tests with high confidence level by the cosmological observations
such as the CMB anisotropy, LSS and {\it etc}.
Thus, it is so that the standard $\Lambda$CDM model still remains well agreement with the majority of cosmological data.

In this paper, we have proposed that the Finsler spacetime with inhomogeneity and anisotropy
could account for the observational formulas of the variation of the fine structure constant $\alpha$.
Based on the Finslerian geodesic equations in the FRW--Randers--Finsler spacetime,
a uniform cosmological redshift was obtained, which is deviated from the one in the FRW--Riemann spacetime.
And the deviation could be revealed generally as one monopole plus dipole function about spacetime locations and directions.
Such a monopole plus dipole function is found to possibly account for the formulas of the $\alpha$ variation.
Thus, the observations of the variation of the fine structure constant from the quasar absorption spectra
could be viewed as a test of the inhomogeneity and anisotropy at large scales.
As was mentioned above, the standard theories could not reasonably explain any $\alpha$ variation
while the FRW--Randers--Finsler spacetime with inhomogeneity and anisotropy could.
This may signal hints for new physics.

\begin{acknowledgments}
We would like to thank Prof. C. J. Zhu and C. G. Huang for some useful discussions. The
work was supported by the NSF of China under Grant No. 10525522,
10875129 and 11075166.
\end{acknowledgments}

\end{document}